\documentclass[10pt]{article}

\usepackage{epsfig}
\usepackage{tabularx}
\usepackage{graphicx} 
\usepackage{color}
\usepackage{xspace}
\usepackage{verbatim}
\usepackage{colortbl}
\usepackage{algorithm}
\usepackage{algorithmic}
\usepackage{amsfonts}
\usepackage{amsmath}
\usepackage{color}
\usepackage{epstopdf}

\usepackage{eso-pic}
\usepackage{flushend}
\usepackage{amssymb}
\usepackage{graphicx}
\usepackage{caption}
\usepackage{subcaption}

\usepackage{hyperref}
\usepackage{bbm}
\usepackage{url}

\def\etal{{\em et al. }}

\newcommand{\E}{\mathrm{E} }
\newcommand{\EFCT}{\mathrm{E[FCT]} }
\newcommand{\EBCT}{\mathrm{E[BCT]} }

\newtheorem{theorem}{Theorem}
\newtheorem{corollary}{Corollary}

\usepackage{geometry}

\date{}

\usepackage{authblk}
\title{Size-based scheduling vs fairness for datacenter flows: a queuing perspective}

\author[1]{James Roberts}
\author[2]{Dario Rossi}
\affil[1]{Independent researcher, France (james.walter.roberts@gmail.com)}
\affil[2]{Huawei Paris Research Center, France (dario.rossi@huawei.com)}

\begin{document}

\maketitle
\begin{abstract}
Contrary to the conclusions of a recent body of work where approximate shortest  remaining processing time first (SRPT) flow scheduling is advocated for datacenter networks,  this paper aims to demonstrate that per-flow fairness remains a preferable objective.  We evaluate abstract queuing models by analysis and simulation to illustrate the non-optimality of SRPT under the reasonable assumptions that datacenter flows occur in \emph{batches} and \emph{bursts} and not, as usually assumed, individually at the instants of a Poisson process.
Results for these models have significant implications for the design of bandwidth sharing strategies for datacenter networks. In particular, we propose a novel  ``virtual fair scheduling'' algorithm  that enforces fairness between batches and is arguably simple enough to be implemented in high speed devices. 
\end{abstract}

\section{Introduction}

To realize low latency in datacenter networks (DCNs) it is now frequently proposed that flow scheduling should be size-based, referring to the well-known response time optimality of the shortest remaining processing time first (SRPT) policy \cite{Schrage1968}. Authors have devised practical algorithms that closely approximate SRPT and demonstrate significant latency reduction compared to max-min fairness, e.g., \cite{Alizadeh2013, Montazeri2018, Mushtaq2019}. However, to demonstrate the advantages of SRPT, the cited papers adopt a simplified traffic model where flows of random size arrive according to a Poisson process. The objective of the present paper is to show that the claimed superiority does not hold under more realistic traffic models and fairness arguably remains a desirable scheduling objective.

While the nature of DCN traffic is imperfectly understood and certainly varies considerably from one instance to another, there emerge from the literature two characteristics that have a significant impact on the performance of flow scheduling. First, the partition/aggregate structure of datacenter applications implies flows do not occur singly but rather in \emph{batches} and the significant performance indicator is not flow completion time (FCT) but rather batch completion time (BCT), the time to complete every flow in the batch. 
Second, flows or batches of flows occur in \emph{bursts}, one batch only beginning after the previous one has completed. This means the arrival process is not independent of the scheduler, a necessary condition for the optimality of SRPT as proved by Schrage \cite{Schrage1968}. 

We investigate the impact of batches and bursts using simple queuing systems modelling a single bottleneck link considered in isolation and exactly realizing SRPT or processor sharing (PS) service disciplines. Analysis and simulation is used to illustrate the impact on the relative performance of SRPT and PS of salient traffic characteristics like the distribution of batch and burst sizes. A notable theoretical contribution is a derivation of the expected BCT of the M$^X$/G/1 preemptive shortest job first (PSJF) queue that closely approximates SRPT. 

The considered models are idealizations but their results have practical implications on the design of  DCN bandwidth sharing mechanisms. In particular, we suggest fair sharing between batches of flows is a desirable pragmatic objective, being simpler to implement than approximate SRPT and having better BCT performance in some practically relevant cases. A further contribution of this paper is to indicate how such sharing might be realized using a novel ``virtual fair scheduling'' algorithm that is arguably simple enough to be implemented in high speed DCN switches.

In the next section we discuss DCN traffic characteristics, identifying the batch and burst structure of flow arrivals. In Sec. \ref{sec:parallel} we consider the M$^X$/G/1 queue under PSJF, SRPT and PS disciplines while in Sec. \ref{sec:successive} we compare SRPT with PS when flows arrive in bursts. The use of the novel virtual fair scheduling algorithm to realize per-batch fair sharing is discussed in Sec. \ref{sec:VFS}.

\section{DCN traffic characteristics}
There is, of course, no general purpose model of datacenter traffic. In this section we seek only to extract from the literature some salient features that need to be taken into account in evaluating the respective performance of size-based scheduling and fairness. 

\subsection{Flow arrivals are not Poisson}
It is notable that, to our knowledge, there are no published results that show that flows in a DCN ever occur as a Poisson process although this is the assumed traffic model in papers advocating size-based scheduling \cite{Alizadeh2013, Montazeri2018, Mushtaq2019}. Results in frequently cited papers, like \cite{Benson2010}, for instance, actually show the contrary while rare studies like \cite{Abad2012} that analyze the arrival process in depth have exhibited self-similarity.  In fact, flows do not occur singly but in \emph{batches}, with multiple simultaneous data transfers proceeding in parallel, and \emph{bursts}, where new batches of flows begin only when the previous batch has completed. The correlation induced by this structure significantly impacts the performance of schedulers.

\subsection{Flows occur in batches}
Chowdhury and Stoica \cite{Chowdhury2012} coined the term \emph{coflow} to describe a collection of flows generated by cluster computing applications. The flows have endpoints in one or more machines and share a common performance goal in that all flows typically need to complete to fulfill that goal. Dogar \etal \cite{Dogar2014} similarly recognized that flows typically occur in batches, notably for web query type applications where a single request is  partitioned among a large set of workers that all respond in a short space of time. Collectively scheduling flows in a coflow has been shown to bring significantly shorter completion times than independent per-flow scheduling  \cite{Chowdhury2014, Dogar2014}.  

While coflows are prevalent in cluster computing, the impact of flow scheduling on job completion time may not be highly significant since data transfer only counts for a very small fraction of this \cite{Ousterhout2015}. Flow scheduling is much more critical for web query-like applications where request processing times are very short and response time depends heavily on network delays. The preponderance of network delays in this context is well-known and indeed is becoming more pronounced as DCNs increasingly adopt remote direct memory access (RDMA) technology, bringing ever smaller processing times, e.g., \cite{Kumar2020}.

Some statistics on query traffic are provided in papers describing the use of memcached in Facebook datacenters \cite{Atikoglu2012, Nishtala2013}. Each user request managed by a web client gives rise to hundreds or thousands of object retrievals from a cluster of cache servers.  Each object is around 1 KB but retrievals are grouped together with an average of 24 objects included in a single cache-client flow. The number of flows in a batch here depends on the number of cache servers and is not highly variable. 

The size of the batch, also known as the incast degree when the flows use a common link~\cite{Alizadeh2010}, may be much larger than that reported in \cite{Nishtala2013}. For instance, Google reports incast degrees that may be measured in thousands for the BigQuery application~\cite{Kumar2020}.

\subsection{Flows occur in bursts}
Flows in a coflow are by definition independent, in the sense that the input of a flow does not depend on the output of another in the same coflow \cite{Chowdhury2014}. Additional correlation in the flow arrival process arises because some cluster computing applications proceed in stages, one stage beginning only when the previous stage is finished. Similarly, a web client will manage user requests sequentially, yielding an arrival process where a new coflow will only occur after the previous one is complete. 

The exact burst structure of flow arrivals depends on the specific application and is not well-understood. However, while there is no general and widely accepted  burst model, it is commonly agreed that this characteristic is clearly present in DCN traffic. Some cluster computing models discussed in \cite{Chowdhury2012}, like ``bulk synchronous parallel'', proceed in supersteps: the coflow of one superstep only begins after a barrier synchronization event. 
Web queries naturally occur in bursts as each response often leads to a new request while a request may itself generate a sequence of dependent coflows \cite{Nishtala2013}. Web clients may also handle a continuous stream of distinct end-users as requests are dispatched by load balancers to idle clients. 

It is worth noting here that the well-known mean completion time optimality of SRPT is, in fact, only proved for arrival sequences that are independent of the service process \cite{Schrage1968}. If a flow can only start some time after a previous flow has ended, it may well be that size-based scheduling is less desirable than other options like fair sharing.

\subsection{Queuing models}
To gain insight and better understand the impact on performance of the above DCN traffic characteristics, we consider two abstract queuing models. The models relate to an isolated network link  (e.g., from the top-of-rack  switch (ToR) to a server hosting web clients) receiving flows that arrive in batches and in bursts. 

The first applies to a link that receives batches of flows, all starting simultaneously at the instants of a Poisson process: the objective is to compare size-based scheduling and fair sharing under different assumptions regarding the flow size and batch size distributions. 
The second model accounts for burst arrivals: in the light of results for the first model and previous work on coflow scheduling~\cite{Chowdhury2014,Dogar2014}, the model considers successive coflows in the burst as a whole, rather than scheduling individual flows.

\section{Batches of flows}
\label{sec:parallel}

We seek to compare the expected batch completion time, E[BCT], for the M$^X$/G/1 system with size-based scheduling and fairness. 
Batches of flows arrive as a Poisson process at rate $\lambda$. The number of flows in a batch $B$ is an independent and identically distributed (i.i.d.) random variable. The individual flow size is also i.i.d. with distribution $F(x)$ and density $f(x)$ and, to avoid unhelpful complications, this distribution has no atoms. The link has unit capacity. Following \cite{Chowdhury2014}, the size of its largest flow is the \emph{batch length} $L$, the number of flows is the \emph{batch width} $B$, and the sum of flow sizes is the \emph{batch size} $S$.
We first consider Preemptive Shortest Job First (PSJF) scheduling, that has similar performance to SRPT while being simpler to analyse. 

\subsection{PSJF}
\begin{figure}[th]
\center
\includegraphics[width=.5\textwidth]{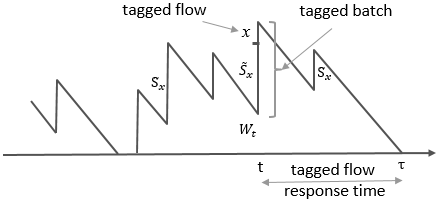}
    
    \caption{Response time of tagged flow as a busy period.}
    \label{fig:busyperiod}
\end{figure}

\begin{figure*}[t]
    \centering
    \begin{subfigure}[b]{0.45\textwidth}
       \includegraphics[width=\textwidth]{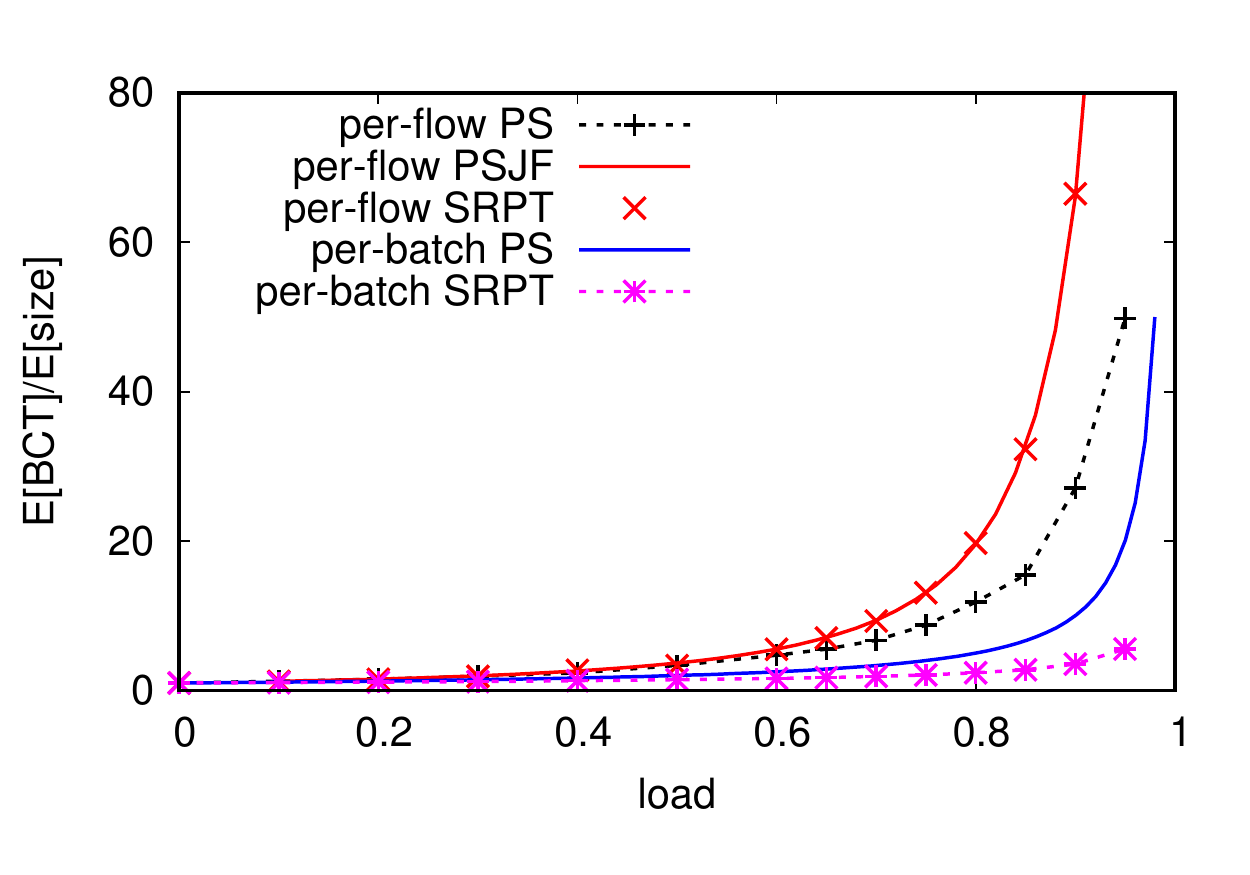}
        \caption{flow CV$^2=0.1$, batch CV$^2=1$} 
        \label{fig:cv2-01-1}
    \end{subfigure}
    \hspace{3em} 
    \begin{subfigure}[b]{0.45\textwidth}
          \includegraphics[width=\textwidth]{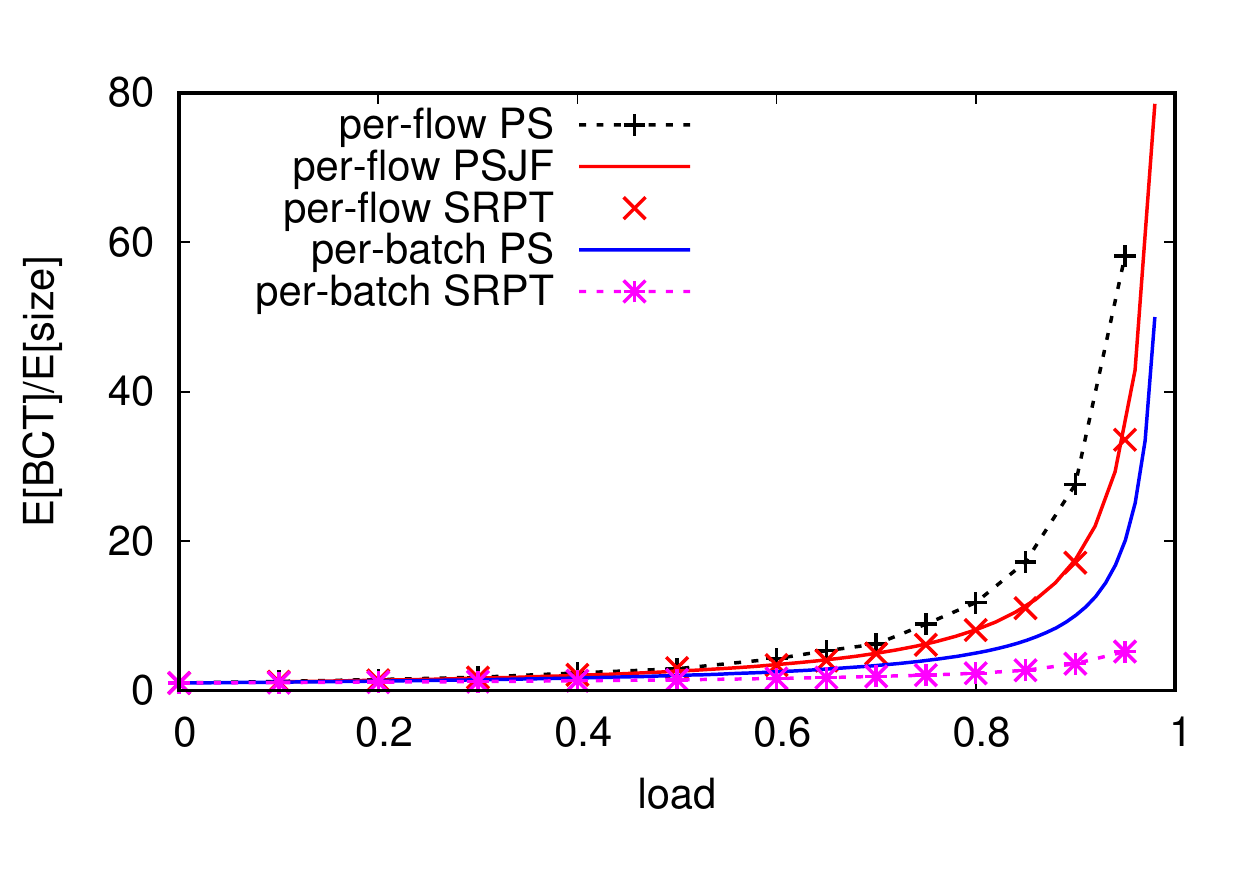}
        \caption{flow CV$^2=10$, batch CV$^2=1$} 
        \label{fig:cv2-10-1}
    \end{subfigure}
     
    \caption{Batch arrivals: normalized expected batch completion time (E[BCT]$/\E[S]$) as a function of the load. Plots generated with geometric batch width, mean $=100$, flow size CV$^2=0.1$ (left) or CV$^2=10$ (right); crosses are from simulations, full lines from analysis.}
    \label{fig:bct}
\end{figure*}

For M/G/1, the performance of PSJF is close to that of the optimal SRPT \cite{Wierman2007}. Here we derive E[BCT] for PSJF with batch arrivals.  Consider a tagged flow of size $x$ and refer to the batch it arrives in as the tagged batch. The tagged flow response time is equal to a certain residual busy period in a work conserving M/G/1 queue where customers are i.i.d. batches of flows of size $<x$ having combined size $S_x$, as illustrated in Fig. \ref{fig:busyperiod}. The residual busy period begins at the tagged batch arrival time, $t$, when the total work in system is equal to the sum of three components: 

\noindent i) $W_t$, the work in system at $t$ due to flows of original size $<x$; 

\noindent ii) $\tilde{S}_x$, the combined size of flows of size $<x$ in the tagged batch; 

\noindent iii) the tagged job itself of size $x$.

\noindent It extends until the tagged flow completes at time $\tau$ and includes the service time of any batches of flows of size $<x$ arriving between $t$ and $\tau$.  Let $m_i(x) = \int_0^x t^i f(t)dt$ be the $i^{th}$ moment of the size of flows of size $<x$ and let $\rho(x)=\lambda \E[B] m_1(x)$ be the system load due to such flows. The proof of the following is given in the appendix. 
\begin{theorem}
\label{th:psjf}
For the M$^X$/G/1 PSJF system, we have,
$$ \EFCT = \int_{x\ge 0} \left(\frac{\E[W_t]+\E[\tilde{S}_x^f]+x}{1 - \rho(x)} \right) f(x)dx, $$
and
$$ \EBCT = \int_{x\ge 0} \left(\frac{\E[W_t]+\E[\tilde{S}_x^b]+x}{1 - \rho(x)} \right) g(x)dx, $$
where
\begin{eqnarray*}
\E[W_t] &=& \frac{\lambda (\E[B(B-1)] m_1(x)^2 + \E[B] m_2(x))}{2(1-\rho(x))},\\
\E[\tilde{S}_x^f] &=& (\E[B^2]/\E[B]-1) m_1(x),\\
\E[\tilde{S}_x^b] &=& \frac{\E[B(B-1)F(x)^{B-2}]}{\E[BF(x)^{B-1}]} m_1(x),
\end{eqnarray*}  
and $g(x) = \E[BF(x)^{B-1}]f(x)$ is the density of the tagged batch length.
\end{theorem}
The following is used later to compute results in  Fig. \ref{fig:bct}.
\begin{corollary}
If $B$ has a geometric distribution of mean $\beta$ we have the following formulas:
$\E[W_t] = (\beta^2m_1(x)^2+\beta m_2(x))/(1-\rho(x))$, $\E[\tilde{S}_x^f] = 2\beta m_1(x)$, $\E[\tilde{S}_x^b]=2 \beta/(1+\beta-\beta F(x))$ and $g(x)=\beta/(1+\beta-\beta F(x))^2$.
\end{corollary}

Numerical experiments with the formulas of Theorem \ref{th:psjf} confirm that PSJF performance is highly sensitive to the flow size distribution with E[BCT] decreasing as the flow size variance increases. The decrease is, however, less marked than the corresponding decrease in E[FCT]. When the variance is high, the main difference between  E[BCT] and E[FCT] comes from the difference between densities $f(x)$ and $g(x)$ since the conditional batch response times are nearly equal. For lightly varying flow sizes, on the other hand, there is also a significant difference between the own-batch terms $\E[\tilde{S}_x^f]$ and $\E[\tilde{S}_x^b]$ that accentuates the negative impact on E[BCT] of size-based scheduling (see Sec. \ref{sec:compare}).

\subsection{SRPT}
To evaluate E[BCT] we can adapt the analysis of M$^X$/G/1 SRPT of Gebrehiwot \etal \cite{Gebrehiwot2017} where an expression for E[FCT] is derived. 
Their analysis conditions the response time of a tagged flow on four different server states at the arrival instant of the tagged batch. These states remain relevant and the only change in the analysis is to account for the composition of the tagged batch that here has length $x$ while in \cite{Gebrehiwot2017} it just contains an arbitrary tagged job of size $x$. This is used in calculating ``the waiting time of a type-$x$ job caused by jobs in its own batch'', denoted $W^b(x)$ in \cite{Gebrehiwot2017}. 

In the notation of the previous section, the expected combined size of the flows of size $<x$ in the tagged batch is $\E[\tilde{S}_x^b]$ given in Theorem \ref{th:psjf} and
$ W^b(x) = \E[\tilde{S}_x^b]/(1-\rho(x))$.
This expression can be substituted in the formulas from \cite[Th 2]{Gebrehiwot2017} to derive an expression for E[BCT]. This expression is not particularly insightful, however, and numerical evaluation is not straightforward. In our evaluation, we have preferred to use simulation, relying on the analytical results for PSJF to provide insight.

\subsection{PS}
Unlike the regular M/G/1 PS queue, the batch arrival M$^X$/G/1 queue with per-flow PS scheduling has no simple and general performance formulas. The integral equation formulation of Kleinrock \etal \cite{Kleinrock1971} can be solved to derive the conditional response time of an arbitrary flow of size $x$ and consequently E[FCT]. Bansal \cite{Bansal2003} provides a computational scheme for a class of flow distributions while Avrachenkov \etal \cite{Avrachenkov2005} have derived  conditional response time asymptotics. In recent work Guillemin and co-authors \cite{Guillemin2021} 
derive the Laplace transform of the batch completion time but only for the M$^X$/M/1 system with geometric batch width. We therefore again rely on simulation to evaluate  E[BCT] for the M$^X$/G/1 PS system.

\begin{figure*}[t]
    \centering
    \begin{subfigure}[b]{0.4\textwidth}
      \includegraphics[width=\textwidth]{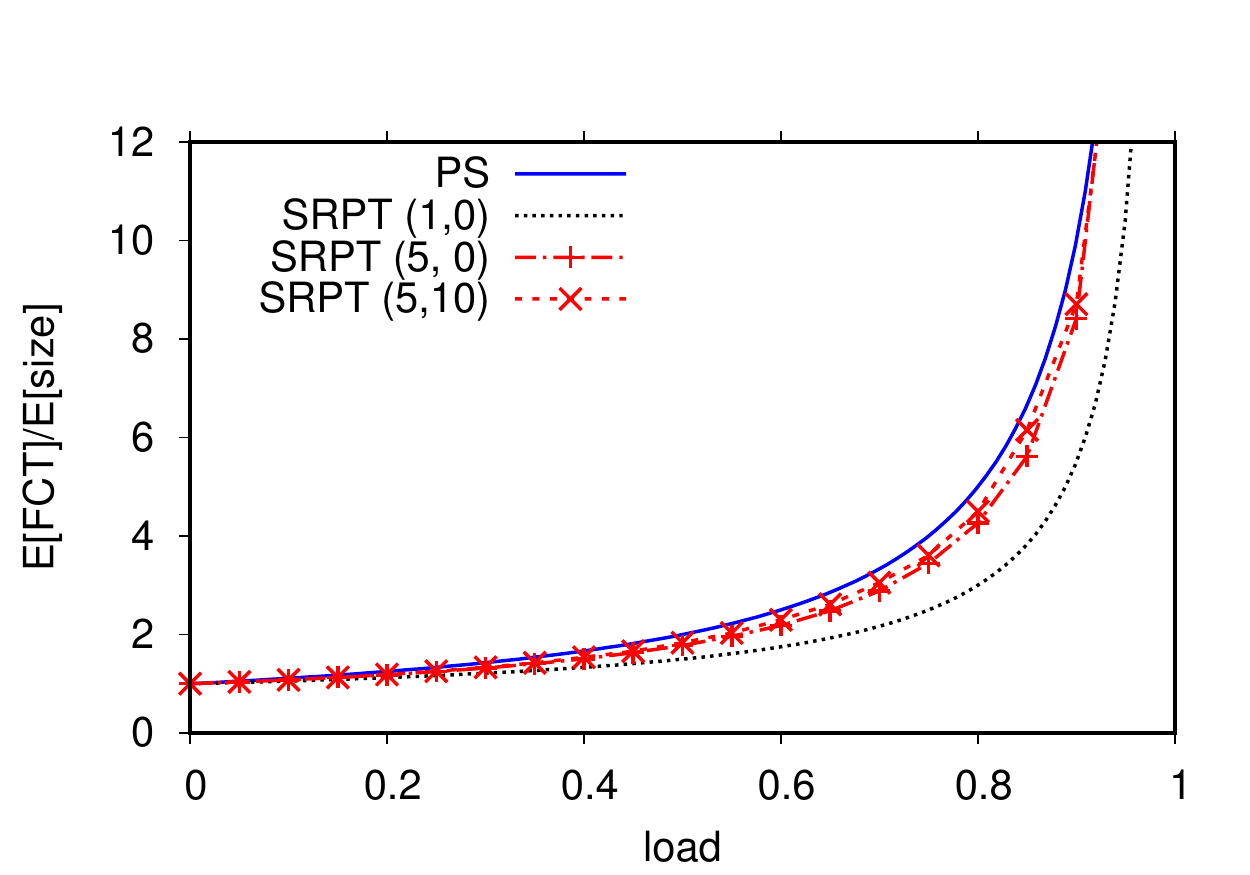}
        \caption{flow size cv$^2$=0} 
        \label{fig:histW6000}
    \end{subfigure}   
 \hspace{3em} 
%
    \begin{subfigure}[b]{0.4\textwidth}
     \includegraphics[width=\textwidth]{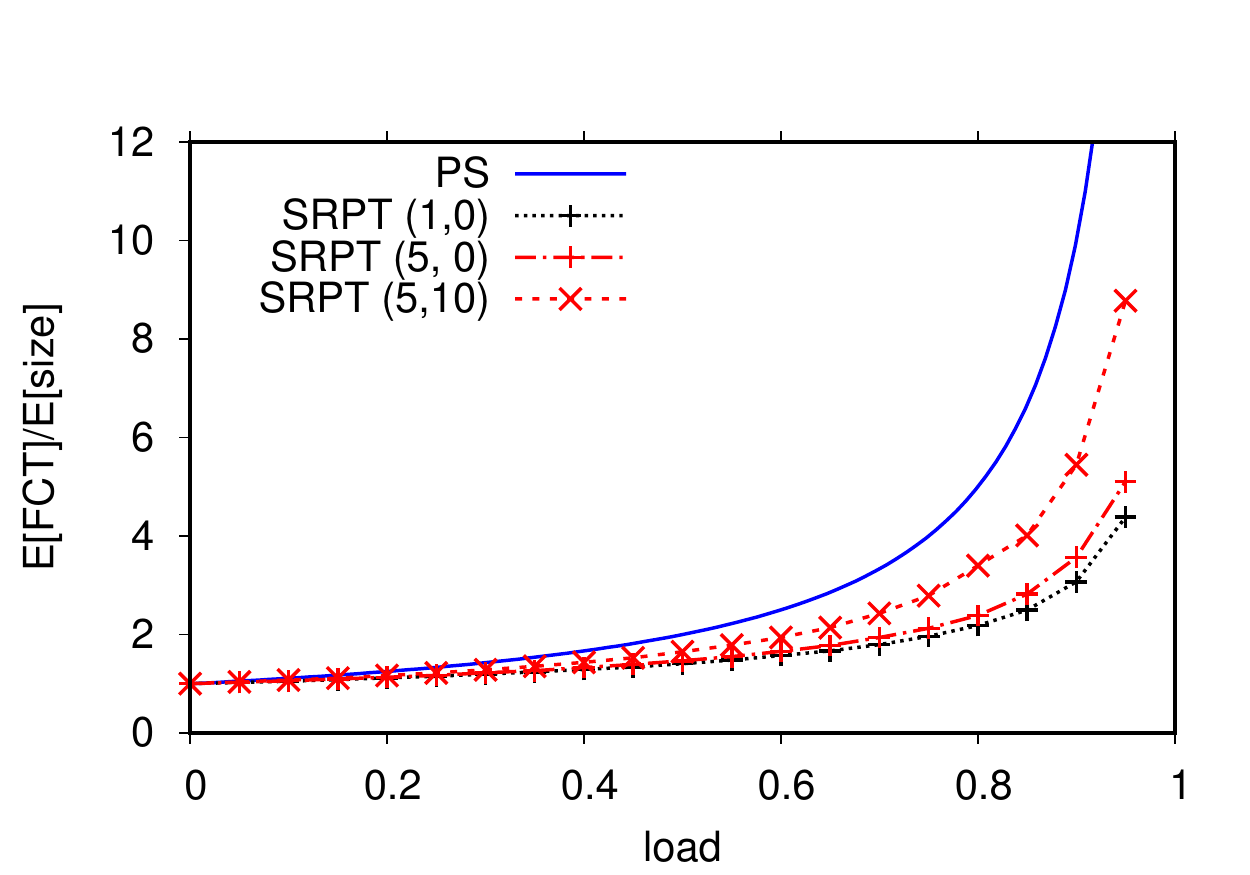}
        \caption{flow size cv$^2$=10} 
        \label{fig:histW9000}
    \end{subfigure}   
 \caption{Partly-open model with homogeneous traffic: normalized expected FCT against load;  flow size is deterministic (left) or Weibull with cv$^2=10$ (right); mean burst length is 1 (denoted `SRPT (1,0)') or 5 with CV$^2=0$ (`SRPT (5,0)') or CV$^2=10$ (`SRPT (5,10)');  exponential inter-flow interval, mean = 1 full rate flow service time.}
\label{fig:partlyFCT}
\end{figure*}

\subsection{Comparative performance}
\label{sec:compare}

We compare the E[BCT] performance of M$^X$/G/1 under PSJF, SRPT and PS scheduling. We further compare per-flow scheduling with per-batch scheduling for SRPT and PS, i.e., where the server capacity is devoted exclusively to the \emph{batch} with the shortest remaining overall size, and where active \emph{batches} share capacity equally, respectively. For these and later comparative results, we ran the simulations long enough to ensure the presented results are accurate at the scale of the figure. 

Fig. \ref{fig:bct} plots E[BCT] against load ($\lambda \E[S]$) for two contrasting unit mean flow size distributions, Weibull with shape parameter 3.5 (squared coefficient of variation, CV$^2\approx 0.1$) and Weibull with shape parameter 0.4 (CV$^2\approx 10$). The number of flows per batch has a geometric distribution of mean 100 (the batch size CV$^2$ is then only 1.035 and 1.004, respectively). These results illustrate the following qualitative properties (confirmed by results for other parameter values):
\begin{itemize}
\item per-flow PSJF has very similar performance to SRPT, the simulation results for SRPT (crosses) being slightly lower than the analytical PSJF results; 
\item per-flow SRPT is preferable to PS when the size distribution has high variance but has higher $\EBCT$ when the variance is small\footnote{Note that SRPT with a deterministic distribution is the same as FCFS and has better performance than PS; this is a singularity not covered by the present assumption that $F(x)$ has no atoms.};
\item per-batch scheduling is significantly more effective than per-flow scheduling;
\item per-batch PS performs less well than per-batch SRPT (that is optimal \cite{Schrage1968}). 
\end{itemize}
Per-flow SRPT has poor E[BCT] performance because the largest flow in a batch naturally becomes a straggler, being neglected in favor of shorter flows in concurrent batches. This phenomenon is accentuated as the size distribution becomes more concentrated about the mean. Per-flow PS also creates stragglers since the largest flow in a batch is eventually alone and in competition with multiple, still active flows in concurrent batches. This phenomenon is worse when the size variance is high since the largest flows then typically compete unfairly for a longer period of time. 
Per-batch scheduling avoids the straggler phenomenon for both SRPT and PS. 

\section{Bursts of flows}
\label{sec:successive}

In this queuing model, single flows\footnote{Note that `flow' here would in fact be all the packets of a batch when scheduling is per-batch, as considered in Sec. \ref{sec:parallel}.} occur in bursts: when a flow completes, it may end the burst or be followed by a new flow after the lapse of an interval of inactivity. This model is a network of two successive service stations, one representing the link that implements either SRPT or PS scheduling, the other an infinite server representing the inactivity interval. Following Schroeder \etal \cite{Schroeder2006} we distinguish a \emph{closed} network model where a fixed number of sources generate successive flows continuously, and a \emph{partly-open} model where bursts arrive as a Poisson process and generate a finite number of successive flows. The case where the ``burst'' is always of size 1 (i.e., Poisson flow arrivals) is termed an \emph{open} model. 

\subsection{Performance of PS and SRPT}
The performance of PS under the partly-open and closed models is well-understood. Average values like E[FCT] are insensitive to the distributions of flow sizes and inactivity intervals that can even be correlated (see \cite[Ch. 3]{Kelly1979}, for example). For the partly-open model, E[FCT]$= 1/(1-\rho)$, as for the open model, for any distribution of the number of flows in a burst. Performance of the closed model depends on the number of sources of different types and can be numerically difficult to compute.

SRPT performance depends sensitively on the distributions of flow size and burst length and we are aware of no useful analytical results for partly-open and closed models. To gain understanding of the comparative performance of SRPT and PS scheduling, we have simulated some specific configurations under a traffic model where flow sizes and inactivity intervals are i.i.d. and mutually independent. The following sections illustrate the impact on SRPT performance of some particular choices of parameter values.

\subsection{Single class, partly-open model}

Figure \ref{fig:partlyFCT} plots the normalized mean FCT against load for a number of homogeneous partly-open model configurations. The interval between flows is exponential with mean equal to the service time of a single flow. We confirmed the observation in \cite{Schroeder2006} that SRPT performance does not depend significantly on the size of the inactivity interval. 
The figure plots normalized E[FCT] for three burst configurations: bursts of exactly 1 flow (i.e., the open model), bursts of exactly 5 flows, and hyper-geometric bursts of 5 flows on average with CV$^2$ of 10. The legend gives the (mean, CV$^2$) combination of the bursts in question. The flow size distribution is deterministic on the left and Weibull with CV$^2 \approx$ 10 (shape parameter .4) on the right.

PS performance is, of course, the same for all configurations. SRPT has consistently lower E[FCT] than PS but the difference decreases as the burst size and variance increase, especially when the flow size CV$^2$ is small. Intuitively, as the burst size gets larger, SRPT switches service between concurrent bursts uniformly at random so that each tends to get a fair share on average. This explains why we can expect SRPT performance to converge to that of PS, i.e., E[FCT]$ \rightarrow 1/(1-\rho)$. SRPT is closer to PS when flow size CV$^2$ is small. Note that this may well be the case in practice when the `flow' in question is actually composed of a large batch of flows, as considered in Sec. \ref{sec:parallel}.

\subsection{Multi-class, partly-open model}

\begin{figure*}[h]
    \centering
    \begin{subfigure}[b]{0.4\textwidth}
      \includegraphics[width=\textwidth]{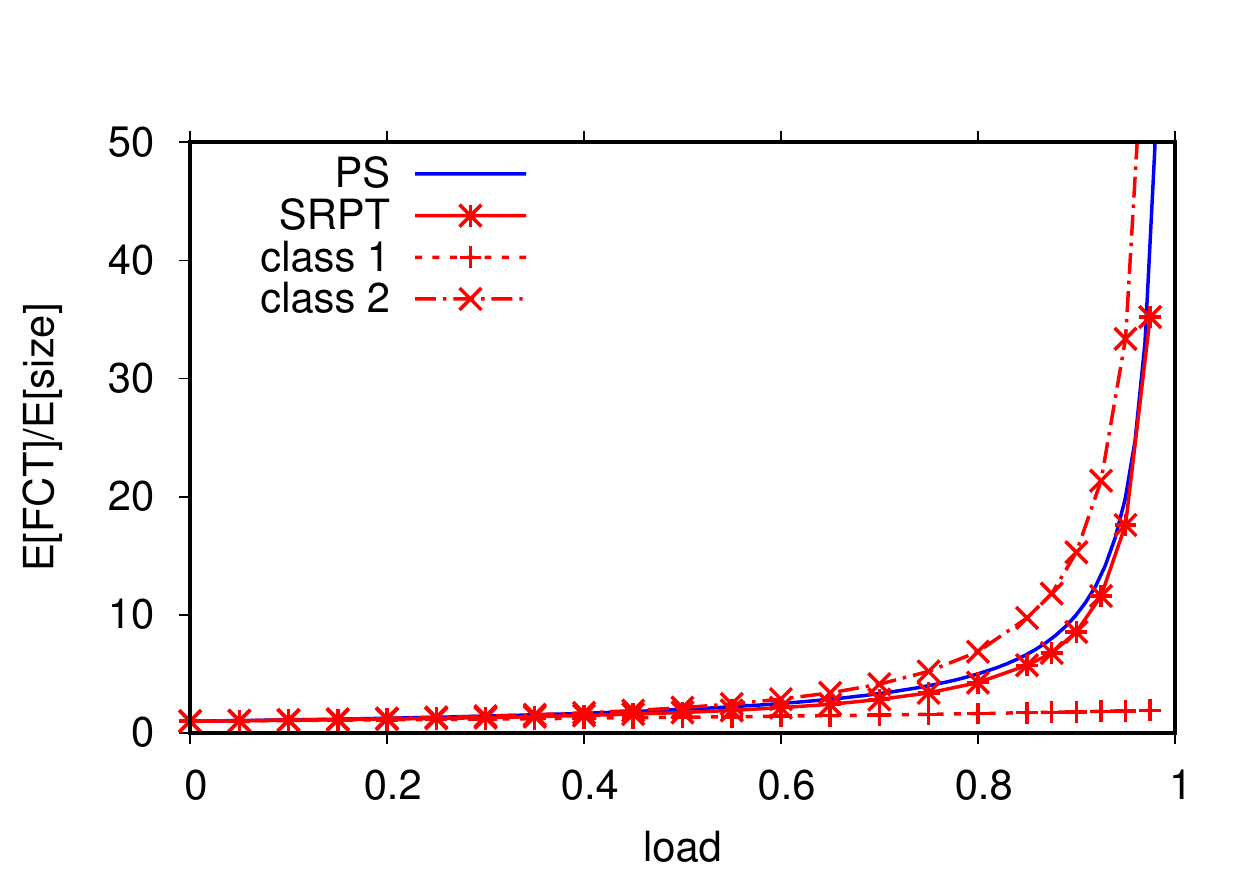}
        \caption{burst mean 5, cv$^2$ 0} 
        \label{fig:2class(5,0)}
    \end{subfigure}
 \hspace{3em} 
    \begin{subfigure}[b]{0.4\textwidth}
     \includegraphics[width=\textwidth]{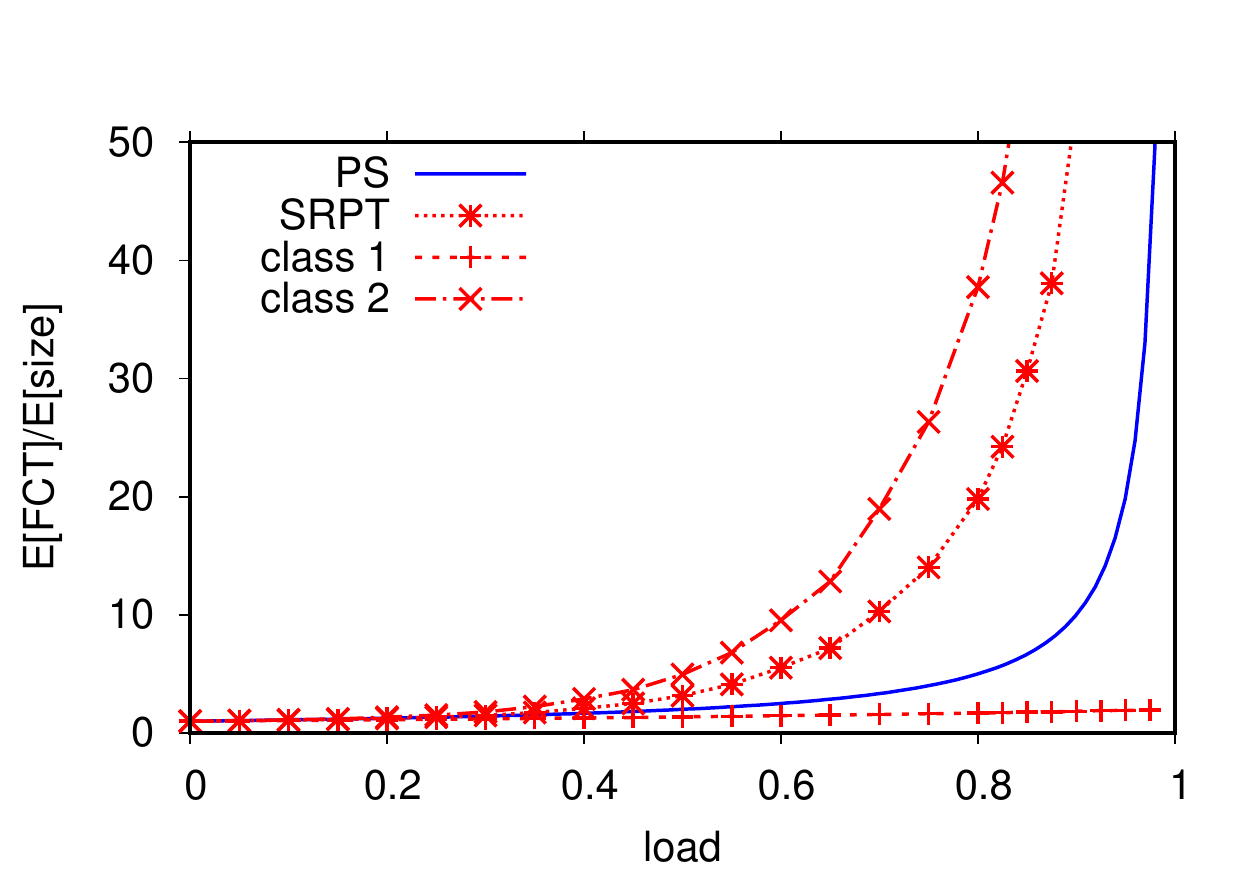}
        \caption{burst mean 5, cv$^2$ 10} 
        \label{fig:2class(5,10)}
    \end{subfigure}   
 \caption{Partly-open model with two classes of traffic: normalized expected FCT against load;  class 1 flows of deterministic size 1 arrive in bursts of mean length 5 and CV$^2=0$ (left) or CV$^2=10$ (right), exponential inter-flow interval, mean = full rate flow service time; class 2 flows of deterministic size 2 arrive singly; each class contributes half the load. }
\label{fig:partlyFCT2cl}
\end{figure*}
\begin{figure*}[h]
    \centering
    \begin{subfigure}[b]{0.4\textwidth}
      \includegraphics[width=\textwidth]{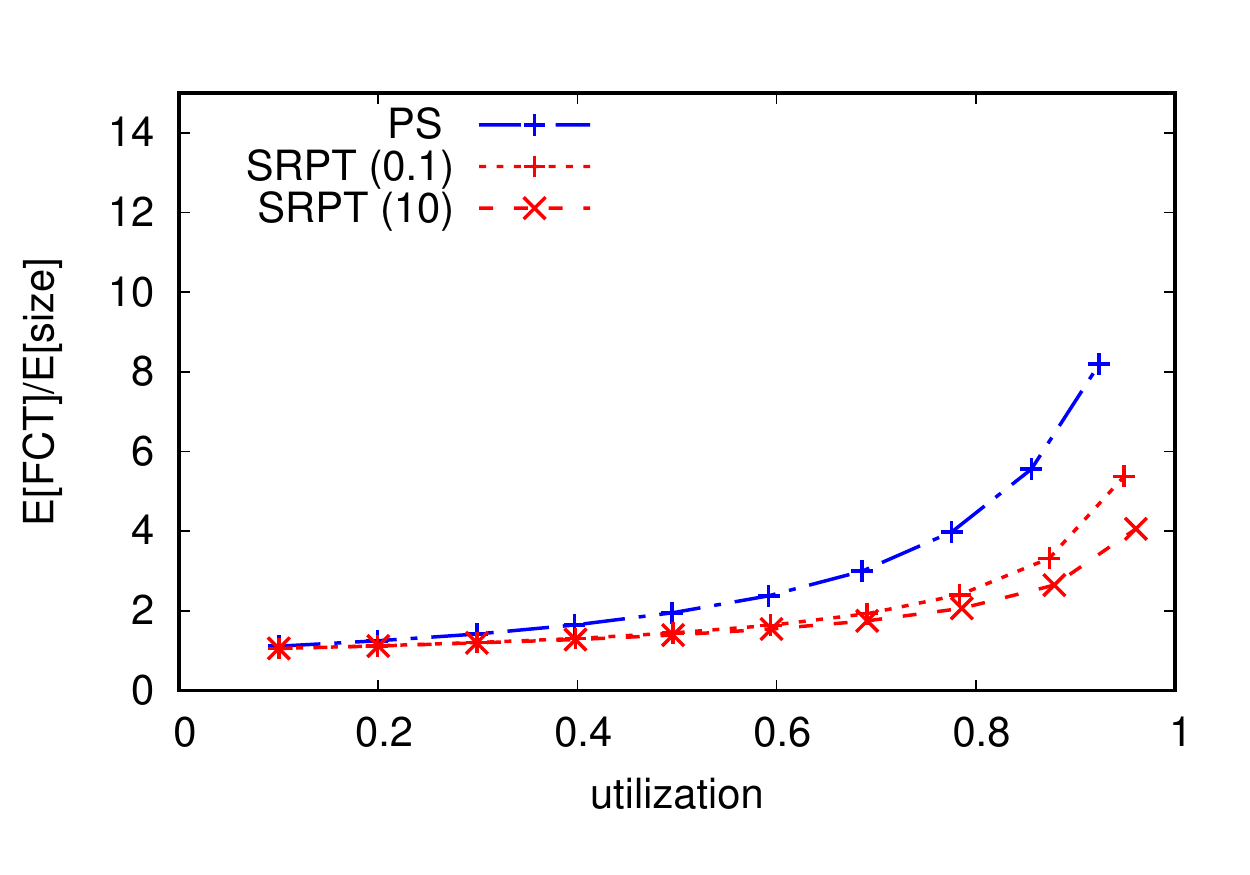}
        \caption{100 homogeneous clients} 
        \label{fig:cl100}
    \end{subfigure}
 \hspace{3em} 
    \begin{subfigure}[b]{0.4\textwidth}
     \includegraphics[width=\textwidth]{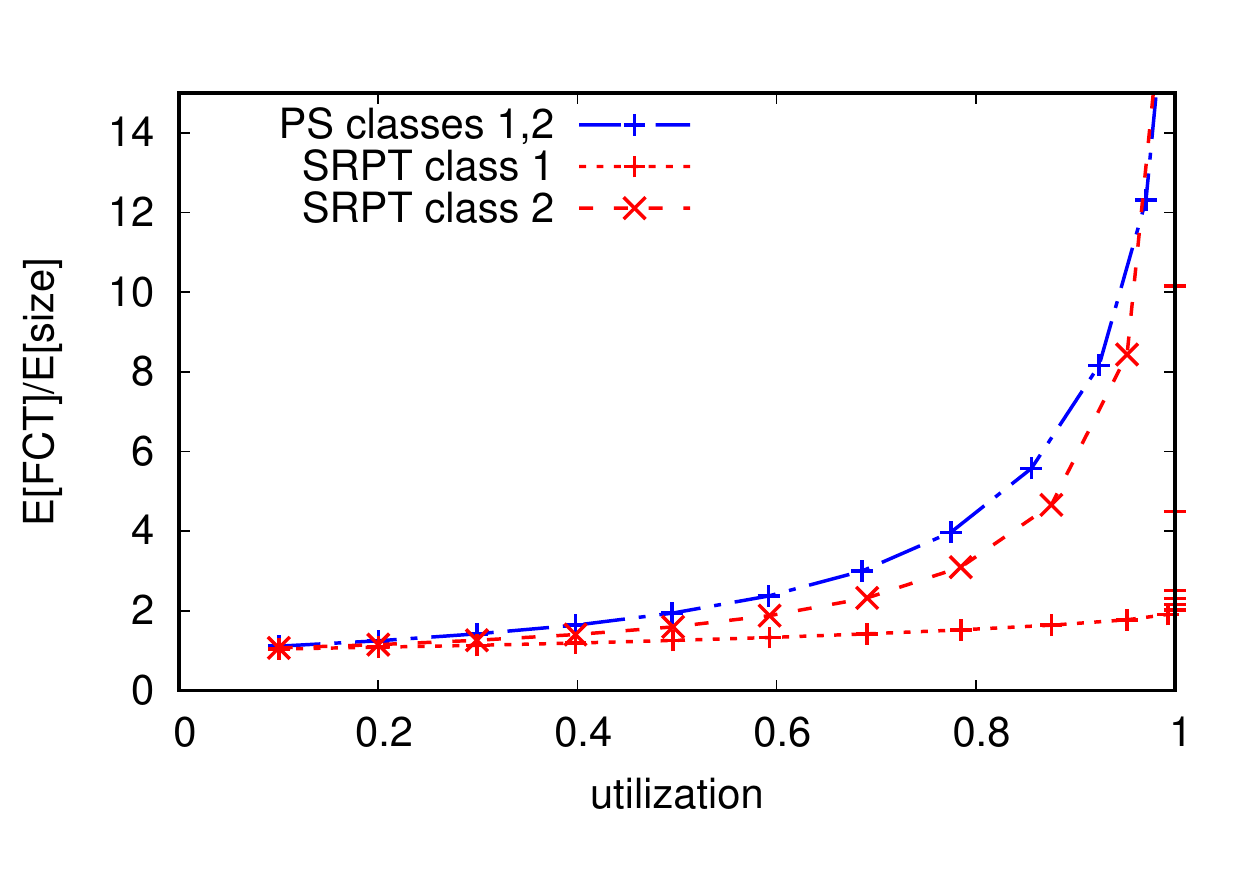}
        \caption{50 class 1 clients, 50 class 2 clients} 
        \label{fig:2class(50,50)}
    \end{subfigure}  
 \caption{Closed model with one (left) and two classes of traffic (right): normalized expected FCT against utilization; left -- 100 clients generate flows with size 1 and CV$^2$ 0.1 or 10; right -- 50 class 1 clients generate flows of size 1 and CV$^2$ 0.1, 50 class 2 clients generate flows of size 2 and CV$^2$ .1; the interval between flows is modulated to generate the range of utilizations keeping the ratio E[flow size]/E[interval] the same for each class.}
\label{fig:closed}
\end{figure*}

When there are multiple classes of flow bursts with distinct characteristics, SRPT can result in worse overall E[FCT] than PS. This is illustrated by the results of Fig. \ref{fig:partlyFCT2cl}.  The figure plots E[FCT] against load when the link receives two classes of flows: class 1 flows are of deterministic size 1 and arrive in bursts of mean 5 and CV$^2$ 0 on the left, and mean 5 and CV$^2$ 10 on the right; class 2 flows are of deterministic size 2 and arrive individually. Each class counts for half the overall load.

The results show that the unfairness of SRPT leads to relatively poor performance for class 2 flows and that this can yield an overall E[FCT] worse than that of PS when the burst variance is high. To understand the negative impact of the class 1 burst size distribution, consider the case where the inactivity interval is very small. The system then behaves like an M/G/1 preemptive priority queue: low priority class 2 response time increases linearly with the second moment of the combined size of flows in a class 1 burst  (e.g., \cite[Sec. 3.2.3]{Wierman2007}).

\subsection{Closed model}

Fig. \ref{fig:closed} presents results for a closed model where a fixed number of `clients' initiate a succession of flows separated by an exponential inactivity interval. The size of the interval is varied to produce a range of link utilizations. 

For the left hand figure, 100 clients emit flows with a Weibull size distribution of CV$^2$ 0.1 or 10.  The results confirm the observation in \cite{Schroeder2006} that completion times are smaller in the closed model than in open and partly-open models. SRPT is better than PS and its E[FCT] is nearly insensitive.

The right hand figure illustrates SRPT unfairness when there are two classes of client. The larger class 2 flows are  penalized though their normalized E[FCT] is only worse than the PS average when utilization tends to 1. Class 2 clients are eventually starved of any service under SRPT while PS converges to a service rate of 1/100 for all clients.

\section{Realizing per-batch fair shares}
\label{sec:VFS}

The evaluation in Sec. \ref{sec:parallel} shows it is advantageous to perform per-batch rather than per-flow scheduling. The purpose of this section is to suggest how per-batch fairness might be realized in the datacenter by means of a novel ``virtual fair scheduling'' algorithm. Per-batch SRPT would perform better than PS under some traffic models  (cf. Sec. \ref{sec:parallel}) but this advantage is not guaranteed when batches of flows occur in bursts (cf. Sec. \ref{sec:successive}). Moreover, per-batch SRPT appears particularly difficult to realize without use of a central scheduler made aware of the size of all flows, as envisaged in work on coflow scheduling (e.g., \cite{Chowdhury2012, Benet2021}). Per-batch fairness, on the other hand, can be realized in a distributed manner, as described below.

\subsection{End-system and switch mechanisms}
To realize per-batch fairness it is necessary to implement mechanisms in both switch and end-systems. Fairness on the ToR-server link would be enforced by a switch mechanism supposed able to recognize packets belonging to a given batch from header fields identifying the client having initiated multiple parallel requests. The end-systems are supposed to implement a transport protocol to manage the individual flows of the batch and efficiently exploit the enforced fair bandwidth share. 

The switch might implement a classical fair queuing scheduler like deficit round robin (DRR) \cite{Shreedar1996} or start-time fair queuing (STFQ) \cite{Goyal1997}. The number of batches that need to be scheduled in the open or partly-open models discussed in Sec. \ref{sec:successive} has a geometric distribution when fairness is enforced and per-batch scheduling is therefore scalable and feasible \cite{Kortebi2005}, e.g., the number of active flows at load .9 is less than 66 with probability .999. 
The scalability of fair queuing has been confirmed in recent work where approximate realizations of the classical scheduling algorithms have been proposed \cite{Sharma2017}, \cite{Goyal2019}. 
To implement approximate fair queuing in high speed datacenter switches remains challenging, however. Our objective here is to suggest a much simpler algorithm, that we call virtual fair scheduling (VFS), is an attractive alternative.

In the next section we present the simplest version of this algorithm that enforces fairness by dropping excess packets. Fair dropping is well-known to be a viable means to enforce per-flow fairness \cite{Pan2003}, \cite{Addanki2018}. End systems are supposed to interpret drops as congestion signals and react appropriately by reducing flow rates, as in classical TCP/IP.  We do not seek here to further define the required transport protocol. We just assume it efficiently distributes the batch fair bandwidth share among flows in a work-conserving manner (e.g., by adapting protocols in \cite{Gao2015} or \cite{Montazeri2018}) .

\subsection{Virtual fair scheduling}
VFS is a significant simplification of the fair dropping algorithm proposed by Addanki \etal \cite{Addanki2018}. While that algorithm arguably enforces \emph{exact} fair shares, it is much too complex to be implemented in a datacenter switch. VFS on the other hand, is less precise but only performs simple comparisons, additions and subtractions.   VFS pseudocode is shown in  Algo. \ref{algo:fdapprox}. A rapid inspection reveals that VFS updates state for only one flow at a time and employs a simple round robin based management of the active flow list. Its simple operation and lean data structure arguably make VFS readily implementable in a modern, high speed DCN switch. 


\begin{algorithm} [t]
\begin{algorithmic}[1]
\STATE 
\COMMENT {On arrival of $n^{th}$ packet at time $t_n$}
\STATE ${credit} \mathrel{+}=C (t_n - t_{n-1}) $  \label{algo:credit}
\STATE input  $pkt$ of $flow(pkt)$
\IF {$flow \in \mathcal{A}$} \label{algo:droporadd2}
\IF {$flow.vq > \theta  $} 
\STATE drop $pkt$
\ELSE \label{algo:newflow2} 
\STATE {$flow.vq \mathrel{+}=pkt.length$} 
\ENDIF
\ELSE \STATE {$\mathcal{A}=\mathcal{A}\cup flow$}
\STATE $flow.vq=packet.length$
\STATE add $flow$ as tail($\mathcal{A}$)
\ENDIF  \label{algo:enddrop2}
\\
\COMMENT {At some epochs $s_m$, eg, $s_n=t_n$}
\STATE $flow=head(\mathcal{A})$  \label{algo:roundrobin}
\IF {${credit} < flow.vq$ } 
\STATE $flow.vq \mathrel{-}={credit}$
\STATE ${credit}=0$
\ELSE \STATE ${credit} \mathrel{-}= flow.vq$
 \STATE $\mathcal{A}=\mathcal{A}\setminus flow$
\ENDIF 
\IF {$\left| \mathcal{A} \right| > 0$}
\STATE $head(\mathcal{A})=next$ \label{algo:enddecrease2}
\ENDIF
\end{algorithmic}
\caption{Virtual fair scheduling}
\label{algo:fdapprox}
\end{algorithm}

As in the fair dropping algorithm of \cite{Addanki2018}, VFS implements a virtual scheduler that determines when flows exceed the fair rate and packets need to be dropped. Packets are actually scheduled in a simple FIFO that is assumed large enough that the probability of saturation is negligibly small given that flow rates are controlled by VFS. 

The algorithm relies on a table of active flows $\mathcal{A}$\footnote{In this section we use `flow' in the general sense of all packets having in common a certain set of header fields; for \emph{batch} fairness, these fields should identify the specific client in the  server receiving the batch.} that records the current occupancy in bytes, $flow.vq$, of a dedicated virtual queue (VQ) for each flow in $\mathcal{A}$. This occupancy is incremented on packet arrival, unless the counter already exceeds a threshold $\theta$ in which case the packet is dropped, and decremented at certain epochs, as explained in the next paragraph. Given the scalability of fair queuing, the flow table $\mathcal{A}$ can be realized efficiently as a hash table, as proposed in \cite{Sharma2017} or \cite{Goyal2019} for example. 

The elements of $\mathcal{A}$ are arranged in a linked list and, at a series of epochs $s_m$ for $m=1,2,\dots$, the VQ of the flow currently at the head of this list is decremented. The decrement is equal to the minimum of a state variable $credit$ and the present value of $flow.vq$. Variable $credit$ gives the amount of service capacity that is so far unallocated to any flow. It is increased at every arrival (line \ref{algo:credit}). If the chosen VQ is reduced to zero, that flow is removed from $\mathcal{A}$ while the residual credit is retained and used at the next epoch $s_{m+1}$ to decrement the next VQ in the linked list.  The flow at the head of the linked list steps in round robin order on each epoch $s_m$. 

This algorithm is fair in the sense that any two flows that are permanently backlogged in the shadow system in some interval receive the same \emph{expected} amount of service capacity. The variance of the service amount each receives depends on the relative frequency of the epochs $s_m$. We experimented with this and found a frequency equivalent to the arrival rate (e.g., decrementing the VQ of the head of $\mathcal{A}$ immediately after incrementing the VQ of the flow of the arriving packet) is sufficient. However, it is not necessary that $s_n=t_n$ for all $n$ and it can be advantageous for practical reasons to set the $s_m$ differently.

\subsection{VFS performance}

\begin{figure}[th]
\center
\includegraphics[width=.6\textwidth]{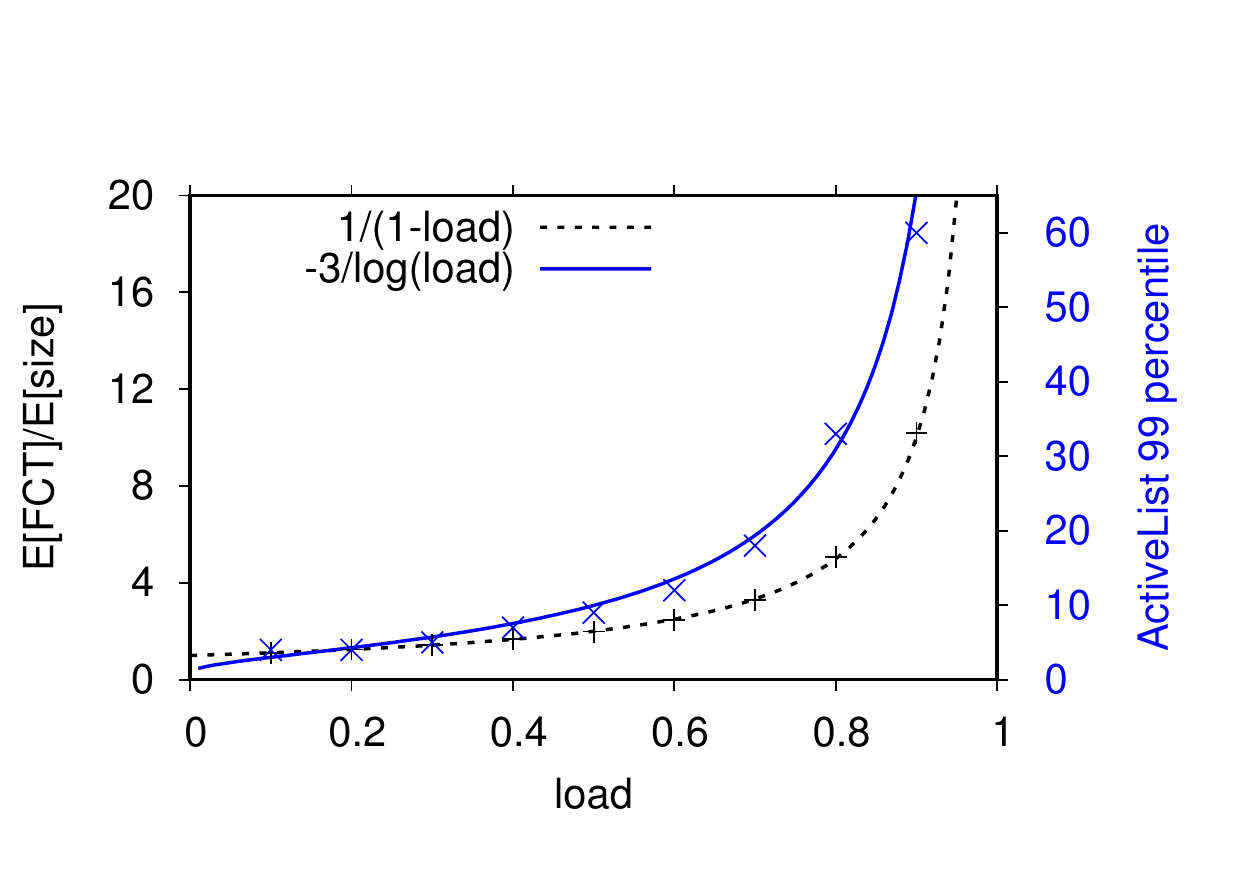}
\caption{Performance of VFS: simulations (crosses) compared to analysis (lines); normalized E[FCT], $1/(1-\rho)$, and 99.9 percentile of active list size, $-3/\log_{10} \rho$.}
\label{fig:fdapprox}
\end{figure}

Assuming the algorithm does enforce fair shares like a PS system, we would expect the normalized E[FCT] at load $\rho$ to be  $1/(1-\rho)$. Moreover, the distribution of the number of flows in the active list would be geometric with a $p$-percentile equal to $\log(1-p/100)/\log(\rho)$. Figure \ref{fig:fdapprox} confirms that this is true for a particular traffic model with VQs decremented following each arrival, i.e., $s_m=t_m$.

The figure plots simulation results as crosses and the analytical PS results as lines, for E[FCT] (left y-axis) and the 99.9 percentile of the distribution of $|\mathcal{A}|$ (right y-axis), against load $\rho$. The traffic is composed of a Poisson process of flows emitting at line rate until 3000 packets have been successfully transmitted together with a Poisson process of single packet flows. The traffic volume in bit/s is 80\% due to the large flows and 20\% due to singletons. The results confirm that the analysis accurately predicts the simulation results for this particular case. We have verified for numerous other configurations that the agreement is true in general. 

\subsection{VFS  enhancements}
The algorithm can be enhanced in several ways. It is possible to ensure low packet latency for flows emitting at a rate less than the fair rate by forwarding their packets via a priority queue \cite{Kortebi2005b, Hoeiland2018}. These packets are identified by the fact that their flow is not already present in $\mathcal{A}$ when they arrive. 

The algorithm can be adapted to provide weighted fair shares. Instead of following the linked list round robin order (line \ref{algo:roundrobin}), one could select the VQ to be decremented with a probability proportional to its weight.

A more significant enhancement would be to replace packet drops as a congestion signal by an explicit notification. We have in mind feeding back to the end systems a measure of the currently realized fair rate. This would allow the end systems to pace packet emissions to avoid loss and delay, thus realizing the goal of lossless networking, as discussed in \cite{Li2019} and \cite{Kumar2020}, for example. The fair rate can be derived by measuring the load due to packets from non-backlogged flows ($flow \notin \mathcal{A}$ on packet arrival), subtracting this from the link capacity $C$, and dividing by $|\mathcal{A}|$.

We have verified the feasibility of these enhancements with small-scale simulations of a single link. It remains, however, to perform a more thorough evaluation and to experimentally validate a possible implementation in a high speed switch.

\section{Conclusion}

By evaluating two abstract queuing systems, we have thrown light on the impact on performance of two particularities of DCN traffic:  (i) flows frequently occur in batches due to the use of the partition/aggregate paradigm, and (ii) flows occur in bursts where a new flow or batch only starts after the previous one has completed. This structure significantly impacts the performance of scheduling algorithms like SRPT and PS and challenges the supposed superiority of the former, previously demonstrated in the literature under an unrealistic Poisson flow arrivals traffic model~\cite{Alizadeh2013, Montazeri2018, Mushtaq2019}.

Analytical and simulation results in Sec.~\ref{sec:parallel} confirm that per-batch scheduling, where the link is devoted exclusively to the batch with shortest remaining processing time (batch SRPT) or is fairly shared between concurrent batches (batch PS), clearly outperforms per-flow scheduling. While per-batch SRPT minimizes BCT for the M$^X$/G/1 queue, per-flow SRPT can be particularly unfavorable and worse than per-flow PS when the flow size distribution has a small variance.

The superiority of (per-batch) SRPT over PS is significantly mitigated, and can even disappear, when accounting for the burst structure of flow arrivals, as shown in Sec.~\ref{sec:successive}. In particular, SRPT is hardly better than PS when the distribution of flow size has low variance and the number of flows in a burst is large. Additionally, SRPT is unfair to larger flows and this can lead to higher overall expected BCT than PS in certain conditions. While SRPT would be preferable for the closed model, where a finite number of clients receive a never-ending sequence of flows, PS performance is not bad and has the advantage of being predictable given the insensitivity of the underlying queuing model.  

In light of the above results, we believe per-batch fairness is a reasonable scheduling objective, especially as this appears much simpler to implement than per-batch SRPT. We have suggested how this objective might be realized using a novel ``virtual fair scheduling'' algorithm, that closely approximates fair queuing while being simple enough to be implemented in a high speed DCN switch.

The presented numerical results clearly cover only a tiny fraction of the parameter space of the considered queuing systems. Moreover, these systems are greatly simplified models of a DCN link and the applied traffic models are idealized representations of the actual flow arrival process. However, we believe the results do usefully highlight the negative impact of the batch and burst structure of DCN traffic on the performance of size-based scheduling algorithms like SRPT. Designers of schedulers and transport protocols need to be aware of this impact and carefully evaluate network performance under a realistic model of its traffic.

\section*{Appendix: M$^X$/G/1 PSJF}
\noindent \emph{Proof of Theorem \ref{th:psjf}}. Applying the standard formula for the expected residual busy period of an M/G/1 of load $\rho(x)$ and initial work  $W_t+\tilde{S}_x+x$  (e.g., \cite{Wierman2007}),
we deduce the mean conditional response time,
$$T(x) = \frac{\E[W_t]+\E[\tilde{S}_x]+x}{1 - \rho(x)}.$$

By PASTA, $\E[W_t]$ is the expected work in an M/G/1 queue with load $\rho(x)$ and customer service time $S_x$ and is given by the Pollaczek-Khinchin formula,
$$\E[W_t] = \frac{\lambda M_2(x)}{2(1-\rho(x))},$$
where $M_2(x)$ is the second moment of $S_x$. 
Introduce the batch size distribution $\pi_b=\Pr[B=b]$ and denote by $X_i$, for $1\le i \le b$, the sizes of flows in a batch of size $b$. Then,
\begin{eqnarray*}
M_2(x)&=& \E[S_x^2] =  \sum_b \E[ (\sum_{1\le i\le b} X_i 1_{X_i<x})^2 ] \pi_b \\
          &=& \E[B(B-1)] m_1(x)^2 + \E[B] m_2(x), 
\end{eqnarray*}
where $m_i(x) = \int_0^x t^i f(t)dt$.

The mean tagged batch size $\E[\tilde{S}_x]$ depends on the nature of the tagged flow. 
To compute E[FCT], the tagged flow is an arbitrary flow of size $x$ and $\tilde{S}_x$, here denoted $\tilde{S}_x^f$, is the combined size of flows of size $<x$ in the tagged batch. The tagged batch size has the distribution, $\pi'_b=b\pi_b/\E[B]$ 
(since the probability the tagged flow is included in a batch of size $b$ is proportional to $b$) 
and, reasoning as above, we have,
\begin{eqnarray*}
 \E[\tilde{S}_x^f]  & = & \sum_b \E[ (\sum_{i\le b-1} X_i 1_{X_i<x}) ] \pi'_b \\
          &=& (\E[B^2]/\E[B]-1) m_1(x). 
\end{eqnarray*} 

For the batch completion time, the tagged flow is the largest flow in its batch and $\tilde{S}_x$, denoted $\tilde{S}_x^b$, is the combined size of all other flows in the batch. Recall that the largest flow in a batch is the batch length $L$ and the combined batch size is denoted $S$. $L$ has distribution $\sum_b F(x)^b \pi_b$ and, therefore, density,
$$g(x)=\sum_b b f(x) F(x)^{b-1} \pi_b= \E[B F(x)^{B-1}]f(x).$$


The expected batch size given $L=x$ and $B=b$ is,
\begin{eqnarray*}
\E[S |L=x \; \mathrm{ and } \; B=b] &=& x + \sum_{i\le b-1} \E[X_i | X_i<x] \\
           &=& x + (b-1) m_1(x) / F(x),
\end{eqnarray*}
while the batch width $B$, conditioned on its length being $x$ has the distribution,
\begin{eqnarray*}
\Pr[B=b | L=x]     &=& \frac{b F(x)^{b-1} \pi_b} {\E[BF(x)^{B-1}]}.
\end{eqnarray*}

We can now derive the expected combined size of flows of size $<x$ in the tagged batch as, 
\begin{eqnarray*}
\E[\tilde{S}_x^b] &=& \sum_b \E[S | L=x \; \mathrm{ and } \; B=b] \Pr[B=b | L=x] -x \\
                      &=& \frac{\E[B(B-1)F(x)^{B-2}]}{\E[BF(x)^{B-1}]} m_1(x).
\end{eqnarray*}


\bibliographystyle{abbrv} 
\bibliography{srpt}

\label{last-page}

\end{document}